# From sleep medicine to medicine during sleep – a clinical perspective

Nitai Bar[1], Jonathan A. Sobel[2], Thomas Penzel[3,4], Yosi Shamay[2], Joachim A. Behar[2*]

(1) Rambam Hospital, Haifa, Israel

(2) Biomedical Engineering Faculty, Technion-Israel Institute of Technology, Haifa, Israel

(3) Interdisciplinary Center of Sleep Medicine, Charite University Medicine Berlin, Chariteplatz 1, 10117 Berlin, Germany

(4) Saratov State University, Saratov, Russian Federation

Corresponding author:

*E-mail: jbehar@technion.ac.il

**Abstract**

Sleep has a profound influence on the physiology of body systems and biological processes. Molecular studies have shown circadian-regulated shifts in protein expression patterns across human tissues, further emphasizing the unique functional, behavioral and pharmacokinetic landscape of sleep. Thus, many pathological processes are also expected to exhibit sleep-specific manifestations. Nevertheless, sleep is seldom utilized for the study, detection and treatment of non-sleep-specific pathologies. Modern advances in biosensor technologies have enabled remote, non-invasive recording of a growing number of physiologic parameters and biomarkers. Sleep is an ideal time frame for the collection of long and clean physiological time series data which can then be analyzed using data-driven algorithms such as deep learning. In this perspective paper, we aim to highlight the potential of sleep as an auspicious time for diagnosis, management and therapy of non-sleep-specific pathologies. We introduce key clinical studies in selected medical fields, which leveraged novel technologies and the advantageous period of sleep to diagnose, monitor and treat pathologies. We then discuss possible opportunities to further harness this new paradigm and modern technologies to explore human health and disease during sleep and to advance the development of novel clinical applications – from sleep medicine to medicine during sleep.

I. Introduction

Sleep, a phenomenon observed among all animals, has long been identified as a vital process for general health and wellbeing[1]. Despite the estimated decrease in average sleep duration in modern society, humans spend approximately one third of their lives sleeping[2]. Sleep research dates back to at least the 18th century and has developed into the medical subspecialty of "sleep medicine", although it still remains on the fringes of medical training[3]. To date, this field focuses mainly on the diagnosis and treatment of sleep-related disorders, as classified in the International Classification of Sleep Disorders (ICSD-3)[4], diagnostic and statistical manual of mental disorders (DMS-5)[5] or international classification of diseases (ICD-11)[6].

*Physiology during sleep*

Normal adult sleep architecture is comprised of cyclic patterns of non-rapid eye movement (NREM) followed by rapid eye movement (REM) sleep, with a single cycle lasting 90-120 minutes. REM percentage and density tend to increase during the night with successive sleep cycles[7]. All major body systems are affected by sleep, including the respiratory, cardiovascular and genitourinary systems, in addition to drug detoxification, metabolic, thermoregulatory, immune and cognitive processes. Additionally, many hormones exhibit a relation to the circadian cycle, the most notable being cortisol, growth hormone and prolactin, with levels directly relating to sleep stages, age and gender. Alongside alterations in hormone levels, the autonomic nervous system is a major driver of the abrupt changes in physiology observed during sleep. Overall, sleep is characterized by a reduction in peripheral sympathetic activity and increase in parasympathetic activity, with a more complex and fluctuating pattern during REM as compared to NREM sleep[8]. This shift in neural homeostasis induces the characteristic changes in sleep physiology of various systems, which, in turn, account for the distinctive manifestations of pathologies during sleep.

Sleep modulates the presentation of respiratory disorders due to decreased chemosensitivity to hypoxia and hypercapnia as well as reduced respiratory muscle tone. This ability to tolerate higher levels of $CO_2$ and lower levels of $O_2$, which is more pronounced during REM sleep compared to non-REM sleep, diminishes work of breathing[9]. As a result, conditions in which there is little tolerance to low work of breathing, such as chronic obstructive pulmonary disease (COPD) and asthma, may be aggravated even in the absence of a respiratory sleep disorder, although concomitant sleep disorders such as obstructive sleep apnea (OSA) are prevalent in these populations[10,11]. Sleep also impacts several cardiac functions and induces decreased heart rate and systolic blood pressure (BP), an effect termed the "dipping phenomenon"[12]. This physiologic shift is postulated to enable some rest to the vascular system, endothelial system and heart mechanics. Impairment of this regulatory effect can mirror cardiac or vascular problems and in some cases may be an early manifestation of a cardiac pathology[8].

*Remote health monitoring and sleep*

The emerging role of portable biosensors in clinical practice has been gaining substantial interest[13]. Portable devices and sensors that can record a variety of biosignals are already the subject of hundreds of clinical trials[14] and notable publishing houses, such as Nature or the Lancet, have opened new journals in the field of Digital Medicine with a high focus on portable medicine. Sleep is an ideal time frame for the collection of long and clean physiological time series data, which can then be analyzed using data-driven algorithms such as deep learning[15,16].

*The knowledge gap*

Little is known about how non-sleep-specific diseases manifest during sleep and the contribution of sleep to their pathophysiology. The unique physiological and biomolecular characteristics of sleep imply that pathologic processes, as well as therapeutic interventions, will also exhibit marked differences when applied during sleep as compared to daytime. Nevertheless, despite the potential advantages, sleep is seldom exploited in clinical practice to diagnose, monitor and treat non-sleep-specific conditions – a paradigm we recently coined 'medicine during sleep'[17]. This perspective article motivates this new paradigm by reviewing existing clinical research that focuses on the diagnosis, treatment and management of non-sleep-specific diseases during sleep.

## II. Diagnosis and management of non-sleep-specific conditions during sleep

<u>Ventilation during sleep</u>

Monitoring arterial $O_2$ saturation and $CO_2$ levels enables evaluation of pulmonary gas exchange. Both can be non-invasively estimated by continuous measurement of peripheral oxygen saturation ($SpO_2$) or transcutaneous $O_2$ ($TcO_2$) and end tidal $CO_2$ ($ETCO_2$) or transcutaneous $PCO_2$ ($TcPCO_2$), respectively. Sleep ventilation assessment is being used in a growing number of respiratory-related clinical scenarios ranging from acute infectious episodes in infants to chronic lung or extrapulmonary disorders affecting respiration, in both inpatient and ambulatory settings[18].

The importance and high prevalence of sleep desaturation in patients with COPD, including those who do not exhibit significant daytime desaturations, have been noted in several studies[19,20]. Due to the variable nature of such desaturations, nocturnal pulse oximetry needs to be implemented for more than one night to detect them at an early stage[21]. Additional findings suggest that desaturations during NREM sleep contribute to brain impairment in COPD[22]. Patients with interstitial lung disease (ILD) also exhibit sleep desaturations and are particularly vulnerable during REM sleep[23]. This may lead to poor sleep quality and interfere with sleep regenerative mechanisms even in the absence of a coexisting sleep disorder. Regular use of nocturnal pulse oximetry to monitor breathing in the management of ILD has been advocated, and the role of $TcPCO_2$ as well as the impact of additional intermittent hypoxia in these already chronically hypoxic patients is a research priority[23]. A high prevalence of sleep hypoxemia, which aggravates pulmonary arterial hypertension, has been reported in patients with precapillary pulmonary hypertension. As daytime $SpO_2$ is not a reliable predictor of sleep hypoxemia[24], nocturnal pulse oximetry has been suggested as part of routine evaluation of this patient population[25]. To expand the diagnostic potential of sleep ventilation assessment, Levy et al. [26] developed the first standardized toolbox for continuous oximetry time series analysis using digital oximetry biomarkers and recently demonstrated the feasibility of COPD diagnosis using nocturnal oximetry[27]. Thus, sleep respiratory monitoring in chronic respiratory conditions such as COPD and ILD, may be the optimal modality for early detection of disease progression and decompensation, enabling timely, disease-modifying intervention.

Nocturnal pulse oximetry has been extensively used in the management of bronchiolitis, although its role in the management of this common infectious condition is a subject of debate. Transient desaturations (even to 70% or less) are commonly observed in infants with bronchiolitis after discharge, but their clinical significance remains unclear[28]. The lack of clinically proven benefits of this practice, alongside concerns regarding unnecessary hospital admissions, prolonged length of stay and additional costs, have resulted in guidelines recommending its limited use[29]. Thus, the complexity in interpreting nocturnal oximetry patterns in infants warrants the development of oximetry digital biomarkers and of their association with defined clinical endpoints, such as readmission, enabling leverage of this important tool to detect sleep desaturation patterns which reflect a more severe condition.

Cardiac monitoring during sleep

Evaluation and monitoring of nocturnal BP is of paramount importance for diagnosis and management of hypertension and its complications. Masked nocturnal hypertension is a well-established phenomenon, referring to patients who only exhibit abnormal hypertensive values overnight[30]. Specific patterns of nocturnal BP are associated with several cardiovascular adverse outcomes, including cardiac remodeling and all-cause mortality, and are recognized as better predictors of these outcomes compared to daytime BP[12]. Furthermore, nocturnal BP monitoring may be important to evaluate before administration of bedtime hypertensive medication in certain patient populations[12]. BP can be non-invasively recorded by intermittent cuff measurements during nighttime, but this technique is usually disturbing and yields only point measurements. Nowadays, several validated devices for non-invasive continuous BP measurement are available[31]. In addition, the pulse-transit-time, a measurement based on the time delay of the pulse wave between two arterial sites, has been shown to accurately reflect dynamics as well as absolute values, to some extent, of BP[32]. Although nocturnal BP is considered a critical tool for diagnosis and risk stratification in hypertensive patients, its current classification is limited to a small number of patterns, based on maximal/minimal values of systolic BP. A more extensive analysis of continuous nocturnal BP measurement could yield more insights into the pathophysiology of this ubiquitous condition.

In clinical practice, there are a few instances where nocturnal ECG is recorded, e.g., Holter ECG, single-lead ECG in polysomnography (PSG) and in cases of implanted pacemakers. In a recent research of ours[33], we demonstrated that over 22% of individuals with undiagnosed atrial fibrillation could be identified by opportunistic data-driven nocturnal screening of the ECG traces recorded in regular PSG studies. Research has also shown that atrial fibrillation events may be more frequent during sleep than daytime[34], but this may not be case for other cardiac abnormalities[35]. Clinical ECG trace analysis guidelines do not define cardiac abnormalities as a function of daytime or nighttime[36]. However, cardiac intervals can change during nighttime. For example, changes in body position can result in ST segment fluctuation[37]. Furthermore, sleep might affect the manifestation of cardiac arrhythmias, such as sinus bradycardia and tachycardia, since the average heart rate during sleep is lower. Thus, there exists a critical gap in our understanding and definitions of the clinical manifestations of sleep versus daytime cardiac abnormalities. In addition, sleep may represent an opportunistic window for clean, continuous ECG recording and for diagnosis. Overall, there is a basis of research suggesting that ECG monitoring during sleep may provide clinical value in certain diagnostic scenarios.

Fetal monitoring during sleep

The fetal heart rate trace is used as a nonstress test to assess fetal well-being. Clinicians and engineers have been researching ways to remotely monitor fetal well-being to better manage complicated pregnancies. Continuous monitoring with cardiotocography (minutes to 1 hour) may be needed to assess the fetal health in the Dawes Redman analysis[38]. The non-invasive fetal ECG is an alternative recording technique that involves placement of ECG electrodes on the maternal belly to measure both maternal and fetal electrical activity of the heart. From the mixture of signals, the fetal ECG and heart rate may be estimated. Yet, although this technique has been around for decades, with incremental improvement over time, it still faces important challenges, including its high sensitivity to noise, which impairs its clinical implementation. In a recent work, Huhn et al.[39] showed that the success rate of obtaining the fetal heart rate trace from a non-invasive fetal ECG was twice higher in night-time than daytime recordings. This is because of the reduced movement-induced noise during sleep in comparison to daytime. This example highlights how monitoring during sleep may offer a cleaner window for robust continuous physiological measurement of important vital signs and enhance care.

Intracranial pressure during sleep

Intracranial pressure (ICP) measurement is considered essential in the diagnosis and management of various neurologic conditions, with a growing corpus of scores and signal analysis methods introduced into clinical decision-making in recent years[40]. ICP values are heavily influenced by posture, with higher values obtained in the horizontal position. Since long and consistent measurements are required for optimal analysis, ICP recording during sleep is considered the "gold standard"[41]. Although primarily utilized in acute settings, such as traumatic brain injury and intracranial hemorrhage or infection, ICP has been implied as a diagnostic tool in chronic conditions including hydrocephalus, idiopathic intracranial hypertension and headaches. In patients with chronic headaches who were examined for suspected isolated cerebrospinal fluid hypertension, abnormal ICP pulsations were associated with nocturnal and postural headaches[42]. Yet, continuous non-invasive ICP measurement technologies still suffer from substantial drawbacks, limiting their clinical implementation and the ability to evaluate their diagnostic potential in chronic conditions[40]. Machine learning algorithms facilitating in-depth analyses, as opposed to the commonly reported mean ICP, are becoming increasingly popular in the analysis of ICP patterns. One prominent example in which machine learning is being adapted into clinical practice is that of B waves, short ICP elevations which are associated with brain dysfunction, with active research studying their significance[43]. Thus, advancements in measurement and analysis technologies may enable leverage of ICP measurements performed during sleep for diagnosis and characterization of chronic neurologic conditions.

Intraocular pressure during sleep

Nowadays, intraocular pressure (IOP), considered the major risk factor of glaucoma progression, can be measured outside office hours in a continuous, invasive or non-invasive fashion[44]. Studies of 24-h IOP curves reveal significant variations in nocturnal IOP patterns between different glaucoma patient subtypes, where both peak pressure, as well as more complex parameters, such as fluctuations, are of value in prediction of disease progression. Furthermore, a differential effect of existing treatment modalities on nocturnal IOP was demonstrated[45]. In normal-tension glaucoma and in treated patients with disease progression despite normal office IOP readings, nocturnal IOP measurements may represent the only window to measure elevated IOP patterns, thus facilitating their diagnosis and supporting management decisions. Yet, the contribution of elevated IOP during sleep to glaucoma pathophysiology remains poorly understood and ongoing research is revealing its extent as well as the characteristics of patient subgroups warranting nocturnal monitoring[46].

Imaging during sleep

From a research point of view, sleep imaging could enhance our understanding of sleep-associated pathologies. To date, imaging during natural sleep is clinically inapplicable due to technologic and organizational restrains, and is thus limited to experimental settings. For example, a study aiming to investigate children with primary nocturnal enuresis (PNE), performed MRI during natural sleep using T2-Relaxation-Under-Spin-Tagging (TRUST), a technique used to estimate brain oxygenation. Compared to controls, children with PNE had evidence of high oxygen consumption during sleep, presumably resulting in greater susceptibility to hypoxia, which is thought to be related to nocturnal enuresis[47]. In a recent study, Ayoub et al.[48] demonstrated the feasibility of visualizing upper airway dynamics continuously and non-invasively during natural sleep by electrical impedance tomography (EIT). This technique could potentially replace the more invasive approach of drug induced sleep endoscopy. Some sleep-related systemic pathologies still require nocturnal invasive monitoring for their diagnosis, such as gastroesophageal reflux disease which requires esophageal pH monitoring[49]. Future development of imaging modalities that can be applied during sleep, particularly modalities which capture physiological processes by continuous recording such as EIT, could theoretically replace more invasive techniques used for diagnosis of such pathologies.

**III. Treatment during sleep**

Non-invasive ventilation during sleep

Nocturnal respiratory therapeutic interventions are being studied in a number of clinical scenarios and indications. Early nocturnal intervention can possibly delay or alter the progression of respiratory failure. For example, nocturnal non-invasive ventilation (NIV) in hypercapnic COPD patients is the mainstay therapy[50], and home-initiation of such therapy is becoming practical and safe[51]. In neuromuscular disease, early initiation of nocturnal NIV has been advocated in patients with sleep hypercapnia and daytime normocapnia[52]. Studies suggest that some patients with cystic fibrosis might also benefit from nocturnal NIV[53]. Nocturnal continuous positive airway pressure (CPAP), a therapeutic modality applied primarily in OSA patients, has been shown to induce systemic physiological changes, including a proven blood-pressure lowering effect on distinct patient subsets with hypertension[54]. Clinical trials have demonstrated additional beneficial effects of nocturnal CPAP in OSA patients, such as improved glycemic control[55]. Other sentinel clinical trials are evaluating nocturnal CPAP in conditions including asthma[56], stroke rehabilitation[57] and cluster headache[58], emphasizing its yet undetermined impact on many systemic conditions. Nocturnal oxygen therapy has also been proposed as home treatment for patients with early-stage COVID-19, potentially diminishing need for subsequent hospitalization[59]. Altogether, interventions to support respiration during sleep are being increasingly considered in a variety of pathologic conditions and recognized as having a broad systemic effect on disease process.

Hemodialysis during sleep

Dialysis treatments pose a clear burden on patients, which has promoted the development of home-treatment modalities. Both home hemodialysis and peritoneal dialysis were introduced over 50 years ago, but current worldwide trends have shifted towards hemodialysis in a medical center setting in the vast majority of patients[60]. However, interest in home therapies, particularly home hemodialysis, is rising, partly due to its potential nocturnal implementation[61]. Nocturnal dialysis regimens have been shown to be safe and possibly superior to conventional therapy in various outcomes, including left ventricular mass, blood pressure medication utilization and quality of life[62]. Although nocturnal regimens are still recommended for select patients only, the Renal Association clinical practice guideline on hemodialysis[63] underscores the benefit of wider home hemodialysis implementation owing to its flexibility, economic benefits and the non-inferiority of nocturnal regimens[64]. Thus, hemodialysis during sleep is expected to become more common as further evidence accumulates.

Circadian medicine and chronotherapy

The circadian clock, the driver of physiological and biological processes that occur in relation to the day/night rhythm, is an important regulator of the sleep-wake cycle. Novel molecular research techniques and data analysis tools have advanced the characterization of the differential transcriptomes, metabolomes and proteomes of human tissues with relation to their circadian cycle[65,66]. Data indicate that >80% of protein-encoding genes, including known molecular targets of existing drugs, show diurnal variation in their expression[67]. These findings suggest that timing of drug delivery, termed chronotherapy, could have a substantial impact on their efficacy. Additionally, it has been shown that several types of cancer cells exhibit dysregulation of their circadian cycle, implying that synchronization of interventions with circadian regulatory components could offer a novel approach to cancer therapy[68]. These recent advancements highlight the importance of the circadian mechanism in health and disease, promoting a "circadian medicine" approach. The discovery and delineation of transcription-translation feedback loops that control circadian oscillations, the subject of a Nobel prize granted in 2017, has drawn additional attention to this field[69].

Pharmacological chronotherapy during sleep

Several drugs demonstrate improved activity and/or a safety profile when administered within a chronotherapeutic schedule that considers timed physiological processes and pharmacokinetics. For instance, night-time-release of prednisone was identified as a beneficial in patients with rheumatoid arthritis, achieving better outcomes by targeting the circadian rhythms of inflammation[70]. In the case of hypertension, nighttime ingestion of angiotensin-converting enzyme inhibitors or angiotensin receptor blockers, was associated with improved efficiency of BP control while reducing dose requirements and adverse effects[71]. Oxitropium bromide, an anticholinergic drug used to treat asthma, showed increased activity when inhaled at night time[72]. Pharmacological intervention during sleep requires a programmable drug administration system that does not affect patient sleep and does not require human intervention. Several solutions have been developed, such as pulsatile drug delivery systems (PDDS)[73], including film-coated tablets[73], implantable pump systems and aerosol drug delivery systems[74]. PDDS can administer drugs into the blood circulation, the gastro-intestinal tract and onto the skin. Thus, sleep-related chronopharmacotherapies are becoming increasingly applied in clinical practice, achieving better outcomes through both increased efficacy and safety profiles. As knowledge regarding the circadian nature of pathologic processes and corresponding drug targets is gleaned, their use is expected to expand further.

Cancer treatment during sleep

Nocturnal administration of chemotherapy was proposed, already in the 1990s, as a means of improving anti-tumor efficacy. However, despite decades of intense research of circadian rhythm in cancer, chronotherapy has had limited impact on the mainstream clinical oncology[75]. This might be explained by its failure to show significant improvement in overall survival in previous large clinical trials[76–78]. However, at the same time, positive effects of nocturnal chemotherapy on toxicity and tolerability have been demonstrated for almost all types of chemotherapeutic drugs, including: 5-fluorouracil, irinotecan, doxorubicin and most platinum-based drugs[79–82]. This clear clinical benefit of safer chronomodulated chemotherapy warrants further investigations with higher doses or with immunotherapy, one of the most promising treatment modalities of recent times. Major limitations of previous chronotherapy trials were the highly generalized patient selection and lack of personalized medicine and genetic profiling. Today, the application of genetic information in personalized cancer medicine has transformed cancer care and is also being studied in the context of chronotherapy, far beyond chemotherapy. Chronomodulated regimens, including nocturnal regimens, of targeted kinase inhibitors, such as alpelisib, lapatinib, sunitinib and erlotinib, which are used for personalized cancer treatments, have shown substantial benefits in multiple animal models as well as in patients[83–85]. For example, nocturnal administration of alpelisib, a PI3K inhibitor recently approved for breast cancer, was associated with better control of glycemia and with a better clinical outcome compared to daytime administration[86]. Several studies have also explored chronomodulated radiotherapy or nocturnal radiodosing in cancer patients. While no clear conclusions regarding efficacy has been reached, due to mixed results of recent trials, it seems that gender and genetic profiles may be determinants of the toxicities and response rates of chronomodulated dosing schemes (e.g., women seem to benefit more from chronoradiotherapy)[87–89]. Consequently, cancer therapy during sleep is a rapidly expanding field, with promising breakthroughs in both tumor response to therapy and adverse effect profiles, and is becoming an important aspect of personalized cancer therapy.

Nocturnal interventions to reduce circadian dyssynchronization

Light pollution at night from external lighting systems in big cities or from smartphone/computer screens, may be responsible for several public health issues as it impairs the circadian clock's normal resynchronization[90]. It has been reported that night exposure to artificial light inhibits the production of melatonin. In addition, a higher risk of obesity, diabetes, cardiovascular disease, depression, sleep disturbances and cancer has been observed in shift workers[91–93]. Intensive care units (ICU), which, in most cases, are under constant light intensity, represent a particularly stressful environment. This is postulated to contribute to ICU delirium, an acute brain dysfunction associated with increased mortality, prolonged ICU and hospital length of stay, and development of post-ICU cognitive impairment[94]. Interventions to reduce this risk and improve patient recovery using lighting systems that mimic the day-night cycle, successfully diminished adverse circadian dyssynchronization-related outcomes[95]. Further investigation of the effect of light exposure (e.g., intensity, light spectrum and rhythms) on human health will support additional circadian-oriented modifications of the healthcare environment.

**Discussion**

This paper aimed to provide a perspective regarding the clinical benefits of nocturnal diagnostic and therapeutic practices for non-sleep-specific diseases. It pointed out the rationale behind this approach, by reviewing nocturnal characteristics of conditions such as respiratory, cardiovascular and ocular pathologies, as well as circadian-related considerations in drug administration and cancer treatment. It discussed advantages of such practices, such as improved patient convenience due to ambulatory diagnosis and treatment options, early detection of pathologies and complications, avoidance of daytime drug side effects and cost reduction. Finally, it presented non-invasive continuous measurement modalities and highlighted sleep as an optimal time for collection of clean and long, continuous physiological time-series data which can be analyzed using machine learning. Essential components of the presented paradigm are summarized in Fig. 1. The prominent nocturnal diagnostic and therapeutic modalities along with selected clinical examples covered in this paper are presented in Panel I.

**Figure 1**

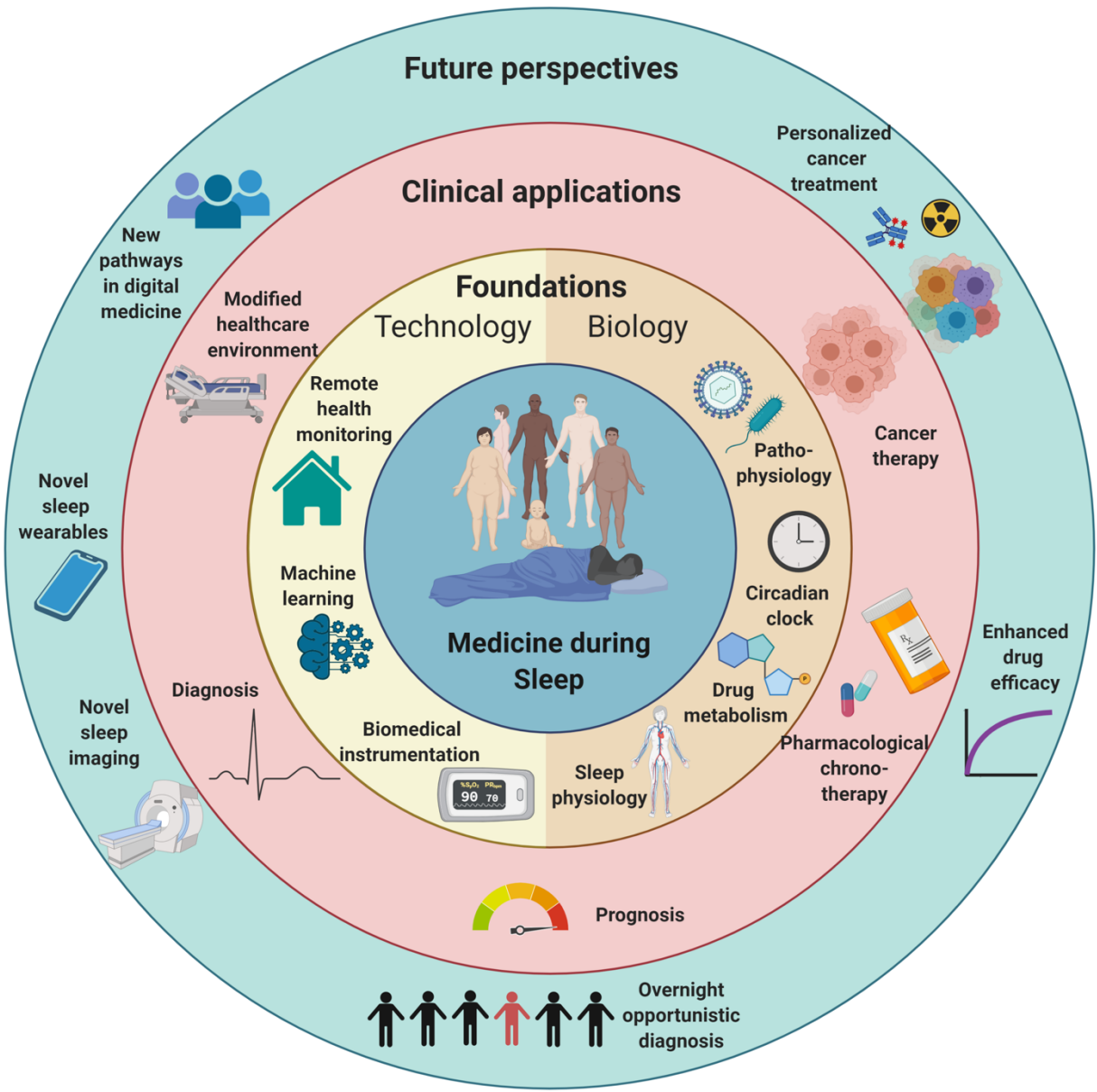

Essential components of the "Medicine during Sleep" paradigm.

**Panel I**
Overview of prominent nocturnal diagnostic and therapeutic modalities and clinical examples highlighting their advantages.

**Respiratory monitoring by $SpO_2$ and $TcPCO_2$**

- Diagnosis of chronic respiratory conditions such as COPD and ILD, and early detection of respiratory decompensation prior to daytime manifestations, possibly representing a preventable cause of further disease aggravation.
- Management of bronchiolitis by characterization of unfavorable sleep respiration patterns. Machine learning algorithms may be able to identify high-risk patients and reduce unnecessary hospital admissions.

**Cardiac monitoring by ECG and BP**

- Diagnosis of masked nocturnal hypertension, characterization of nocturnal BP patterns. Specific nocturnal BP patterns are associated with higher risk for adverse outcomes, and contribute to the choice of medical management of hypertension.
- Diagnosis of arrhythmias manifesting during sleep prior to development of symptoms. Machine learning algorithms may be harnessed to analyze recordings and recognize patterns.

**Ocular monitoring by IOP**

- Diagnosis of glaucoma subtypes in which elevated IOP tends to manifest exclusively during sleep.
- Evaluation of response to therapy in patients which progress despite normal office IOP readings.

**Non-invasive ventilation**

- Early initiation of nocturnal non-invasive ventilation may delay disease progression and extensive need for ventilation in conditions such as chronic lung disease, neuromuscular disease and cystic fibrosis.
- Potential non-respiratory benefits of CPAP include decrease in hypertension and better glycemic control in OSA patients. CPAP is being considered for additional clinical scenarios due to its potential systemic advantages.

**Chronopharmacotherapy**

- In rheumatoid arthritis, nocturnal release of prednisone aims to target circadian inflammatory processes.
- In hypertension, nocturnal administration of ACE inhibitors is associated with improved BP control, reduced doses and fewer side effects.

**Cancer therapy**

- Several chemotherapeutic drugs including 5-fluorouracil, irinotecan, doxorubicin and most platinum-based drugs exhibit reduced toxicity when applied at nighttime.
- Chrono-modulation of chemotherapy, radiotherapy and immunotherapy is expected to become a cornerstone of personalized cancer therapy.

The conceptual foundations of this paradigm have been around for a long time. An excellent example is the work by Verrier et al[8], in which nocturnal cardiovascular physiology and its significance are described in depth. However, clinical applications are still limited. Recent advances in sensor technology and machine learning, as well as discoveries in circadian biology are leading to a growing number of nocturnal clinical applications, albeit still sporadic. With the ongoing efforts to develop more personalized and remote medicine, the time has come to include medicine during sleep as a clinical pathway for diagnostic and therapeutic applications.

Several challenges must be considered in the clinical implementation of this paradigm. Collection and analysis of raw continuous data from non-invasive sensors is relatively new to many medical fields, more so during sleep. As with any newly introduced modality, vigorous clinical groundwork will be required to establish accurate reference ranges and reliable clinical standards. Measurements collected during sleep are likely to differ from daytime measurements, possibly requiring revision of some diagnostic criteria, definitions of significant pathological findings and management of incidental findings. Finally, the introduction of novel sensor technologies specifically designed for medicine during sleep, may enable the realization of this discipline, such as for diagnosis of cardiac pathologies by means of nocturnal BP measurements or ECG.

Another significant challenge arises from the interventional aspect of this paradigm. While non-invasive measurements can be quickly tested in humans, therapeutic interventions require substantial pre-clinical data in animal models. Research of circadian interventions is particularly complicated, since most common pre-clinical models are based on rodents which are nocturnal, with a very different circadian clock than humans. Therefore, effective pre-clinical testing of such interventions will require use of a non-nocturnal mammalian model (e.g., dogs) or implementation of genetic tools to explore the human circadian clock in transgenic or knock-in mice that recapitulate human genetics.

Robust measurement and analysis tools will enable further study of the association between simultaneously measured signals, such as IOP, ICP, cardiovascular and respiratory biomarkers. Deciphering such complex interactions may contribute to a better understanding of pathophysiological processes and their optimal management. Glaucoma is one promising example, where such associations might yield further pathophysiological insights and enable the elaboration of a better diagnostic standard and patient-tailored treatment.

Broadening our knowledge of pathophysiological patterns during sleep may enable harnessing of machine learning algorithms to support screening strategies in selected populations. For example, opportunistic nocturnal screening for hypertension, atrial fibrillation[33] or COPD[27], may be performed in high risk patient populations.

In cancer treatment, chrono-modulation is expanding and is being increasingly explored for novel therapeutic modalities. Studies of the tumor microenvironment have revealed that both CD4 and CD8-T cell levels are correlated with core clock molecules[96], suggesting that chrono-immunotherapy may represent a promising option for future cancer treatment, as well as for other indications[97].

Sleep specialists will play a critical role in the advancement and clinical implementation of sleep-centered diagnostic and therapeutic approaches. Exploitation of this paradigm will require thorough knowledge of sleep physiology and circadian medicine, with emphasis on multidisciplinary training which will broaden sleep diagnosis and treatment beyond current practice. As sleep diagnosis and treatment modalities become a central part of personalized medicine, sleep and circadian medicine will receive more attention in general medical training and within various medical specialties.

**Acknowledgements:**
We are grateful to the Placide Nicod foundation for their financial support (J.S.). TP was partially supported by RF Government grant № 075-15-2019-1885.
**Author contributions:**
All authors have contributed to the writing of this manuscript.
**Competing interest statement:**
All authors have declared no conflict of interest.